\documentstyle[aps,preprint,tighten,floats,epsf,rotate]{revtex}

\begin{document}
\draft

\preprint{\vbox{\it 
                        \null\hfill\rm    IP-BBSR/98-9, March' 98}\\\\}
%
\title{Observing Correlated Production of Defects and Antidefects in
Liquid Crystals}


\author{Sanatan Digal \footnote{e-mail: digal@iopb.res.in},
 Rajarshi Ray \footnote{rajarshi@iopb.res.in}, and Ajit M. 
 Srivastava \footnote{ajit@iopb.res.in}}
\address{Institute of Physics, Sachivalaya Marg, Bhubaneswar 751005, 
India}
%
%
\maketitle
\widetext
\parshape=1 0.75in 5.5in
\begin{abstract}
 We present observations of strength one defects and antidefects
formed in isotropic-nematic phase transition in a thin layer of
nematic liquid crystals, using a cross-polarizer setup. We measure 
the widths of the distributions of {\it net} winding number in small
regions, and determine the exponent characterizing the 
correlation between defects and antidefects to be 0.26$\pm$0.11, 
in very good agreement with the value 1/4 predicted by the 
Kibble mechanism for defect production. We also describe a novel 
technique to determine the director distribution in observations
of defect networks. 
\end{abstract}
\vskip 0.125 in
\parshape=1 -.75in 5.5in
\narrowtext
 
 Study of topological defects is a highly interdisciplinary area
in physics. This has led to a valuable 
interplay of ideas from different branches of physics. For
example, the first theory of formation of topological defects,
formulated by Kibble \cite{kbl} in the context of the early Universe,
found experimental verification of some of its aspects in 
certain condensed matter systems \cite{lc1,lc2,he,zrk}. e.g.  the 
prediction of average defect density, see refs.\onlinecite{lc2,he}.
In this paper, we  present experimental determination of
correlation between defects and antidefects. These results are
important as they present first (to our knowledge) experimental
measurement of defect-antidefect correlations. Also, they are of
importance for theories of defect formation, as they lead to direct 
verification of a prediction of the Kibble mechanism, which is 
qualitatively different from the prediction of defect density.
 
 In the Kibble mechanism, defects form due to a domain structure arising 
in a phase transition. For example, in a spontaneous symmetry breaking 
transition of a U(1) symmetry, with the order parameter being
an angle $\theta$ between 0 and $2\pi$, the order parameter space is a 
circle $S^1$. Domains are characterized by roughly uniform value 
of $\theta$ which varies randomly from one domain to another, (and 
varies with least gradient in between adjacent domains). There are string 
defects here characterized by non-zero winding of $\theta$ around the 
string. By considering the probability of getting a winding around a 
junction of three domains, it is easy to show \cite{lc2} that the 
probability of vortex formation per domain, in 2 space dimensions, is 
equal to 1/4.  

 Consider now a vortex formed at the 
junction of three domains. Given this, the probability of formation 
of an antivortex in the neighboring region increases since part 
of the (anti)winding of $\theta$ is already present, and one only needs 
to have right $\theta$ value, say, in a fourth domain adjacent to the 
other three domains. This conclusion, about 
certain correlation between formation of a defect and an antidefect, is 
valid for other types of defects as well \cite{skrm}. The effect 
of this correlation is the following \cite{vv}. Consider a two
dimensional region $\Omega$ whose area is $A$ and  whose perimeter 
{\it L} goes through $L/\xi$ number of elementary domains (where $\xi$ is 
the domain size). As $\theta$ varies randomly from one domain to another, 
one essentially is dealing with a random walk problem with the 
average step size for $\theta$ being $\pi/2$ (largest step is $\pi$ 
and smallest is zero). Thus, the net winding number of $\theta$ around 
{\it L} will be distributed about zero with a typical width given by 
$\sigma = {1 \over 4} \sqrt{L \over \xi}$, implying that $\sigma \propto 
A^{1/4}$. Assuming roughly uniform defect density, we get $\sigma 
\propto N^{1/4}$ (where $N$ is the total number of defects in the 
region $\Omega$), which reflects the correlation 
between the production of defect and antidefect. In the absence of any 
correlations, net defect number will not be as suppressed, and will 
follow Poisson distribution with $\sigma \sim \sqrt{N}$. In general
one may write the following scaling relation for $\sigma$,

\begin{equation}
\sigma ~ = ~ C ~ N^\nu
\end{equation}

 The exponent $\nu$ will be 1/2 for the uncorrelated case.  As we show 
below, our experimental results give $\nu = 0.26 \pm 0.11$ which is in 
very good agreement with the predicted value of 1/4 from the Kibble 
mechanism, and reflects the correlated nature of defects and antidefects. 
To get $C$ as predicted by the Kibble mechanism, 
we take the elementary domains to be equilateral triangles. Let us 
also assume, for simplicity, that the two dimensional region $\Omega$ 
under observation is a square with area {\it A} = $(L/4)^2$. With the 
probability of defects per domain being 1/4, we get $N = L^2/(16 \sqrt{3} 
\xi^2)$. With $\sigma = {1 \over 4} \sqrt{L \over \xi}$, one can rewrite 
$\sigma$ as $\sigma = (3^{1/8}/2) N^{1/4}$ implying value of $C$ in 
Eqn.(1) to be $3^{1/8}/2 = 0.57$. If the elementary domains are squares, 
then we get $C = 0.71$. (One may have domains of 
various shapes and sizes in a local region. It is not known what should 
be the correct distribution of domains for the Kibble mechanism. Thus, 
as usual, we simply take all domains to be roughly similar.) 

 For uniaxial nematic liquid crystals (NLC) the orientation 
of the order parameter in the nematic phase is given by a unit vector 
(with opposite directions identified) called as the director. The 
order parameter space is $RP^2 (\equiv S^2/Z_2)$, which allows
for string defects with 1/2 winding.
Due to birefringence of NLC, when the liquid crystal sample is placed 
between a cross polarizer setup, then regions where director is either 
parallel, or perpendicular, to the electric field ${\vec E}$, the 
polarization is maintained resulting in a dark brush. At other regions, 
the polarization changes through the sample, resulting in a bright 
region. This implies that for a defect of strength $s$, one will observe 
$4s$ number of dark brushes \cite{chndr}. If the cross-polarizer setup 
is rotated then brushes will rotate in the same (opposite) direction for 
positive (negative) windings. Equivalently, if the sample is rotated 
between fixed cross polarizers, then brushes do not rotate for +1 winding
while they rotate in the same direction (with twice the angle of 
rotation of the sample) for $-$1 winding. We use this method 
to determine the windings.  

  We now describe our experiment.  We observed isotropic-nematic (I-N) 
transition in a tiny droplet (size $\sim$ 2-3 mm) of NLC 
$4'$-Pentyl-4-biphenyl-carbonitrile (98\% pure, purchased from Aldrich 
Chem. Co., Milwaukee, USA). The sample was placed on a clean, untreated 
glass slide and was heated using an ordinary lamp. The I-N transition 
temperature is about 35.3 $^0$C.  Our setup allowed the possibility of 
slow heating, and cooling, by changing the distance of the lamp from the 
sample. (This part was same as in ref.\onlinecite{lc2}.) We 
observed the defect production very close to the transition temperature
(in some cases we had some isotropic bubbles co-existing with the nematic 
layer containing defects). We used a Leica, DMRM microscope with 20x 
objective, a CCD camera, and a cross-polarizer setup for the observations.
Phase transition process was recorded on a standard video cassette 
recorder.  The images were photographed directly from a television 
monitor by replaying the cassette.

 The I-N transition is of first order. We find that when the transition 
proceeds via nucleation of bubbles, one primarily gets  long horizontal 
strings which are not suitable for our analysis.  We selected those 
events where the transition seems to occur uniformly in a thin layer near 
the top of the droplet (possibly due to faster cooling from contact with 
air).  Depth of field of our microscope was about 20 microns. All the 
defects in the field of view were well focussed suggesting that they 
formed in a thin layer, especially since typical inter-defect
separation was about 30 - 40 microns. (For us, the only thing relevant 
is that the layer be effectively two dimensional over distances of
order of typical inter-defect separation).  Also, the 
transition happened over the entire observation region roughly uniformly, 
suggesting that a process like spinodal decomposition may have
been responsible for the transition. This resulted in a distribution
of strength one defects  as shown in the photographs in Fig.1. Points
from which four dark brushes emanate correspond to defects of strength 
$\pm$ 1. Due to resolution limitation, or due to rapidly evolving 
director configuration, the crossings here do not appear as point like. 
It is practically impossible to use the technique of rotation of brushes
to identify every winding in situations such as shown in Fig.1 
due to very small inter-defect separation (resulting form high defect
density), as well as due to the fact that transition proceeded rather 
fast leading to extremely rapid evolution of defect distribution.

 We needed to analyze situations of very high defect density (as in 
Fig.1) due to requirement of large statistics. We have developed a
particular technique for determining individual windings of
defects in situations like Fig.1 where one only needs to determine
the winding of one of the defects by rotation in cross polarizer setup.
Windings of the rest of the defects can then be determined using 
topological arguments. We now explain this procedure using pictures
shown in Fig.2. 

 Fig.2b shows the situation where the sample is rotated in a 
clockwise manner, compared to the situation shown in Fig.2a (as can be 
seen clearly by noticing defect patterns). We first determine winding of 
one of the defects, marked by a $*$ in Fig.2a. By noting the rotation of 
brushes in Fig.2b, we determine that it is a defect ($v$, marked by 
arrow) with +1 winding (as the brushes do not rotate for this defect, 
note, here we are rotating the sample). Now, one of the brushes emanating 
from this defect is assumed to correspond to director being parallel to 
${\vec E}$ with $\theta$ assumed to be zero. Winding = +1 
then implies that the next brush, going 
clockwise around the defect, should correspond to $\theta = 3\pi/2$ with 
the director perpendicular to ${\vec E}$. The next two brushes will then 
correspond to $\theta = \pi$, and $\pi/2$ respectively. We now denote 
the quadrant on the circle $S^1$, between $\theta = 0$ to $3\pi/2$ 
by number 1, the quadrant, between $\theta = 3\pi/2$ to $\pi$ 
by number 2, and similarly, other two quadrants (going clockwise) by 
numbers 3 and 4. This allows us to write numbers 1 to 4 in between the 
brushes. 

  Using continuity of the order parameter outside the location of
defects, one can easily see that $\theta$ in between any two brushes 
remains in the same quadrant irrespective of the size and the shape 
of that region. This is because a change of quadrant should happen only 
across a dark brush. Around any defect, if we know the quadrant 
numbering of any two adjacent regions between the brushes, then the 
quadrant numbers for the remaining two regions are assigned assuming the 
same sense for the winding as determined by the first two quadrant 
numbers. (This is when the crossing corresponds to a defect. Occasionally
there are situations where one is not able to resolve whether two 
brushes are just very close, or there is a crossing there. In such cases,
if a wrong choice is made, then one finds a conflict in quadrant 
assignment when approaching from different directions.) 
Using these simple rules, we complete quadrant assignment in the
picture in Fig.2b, and determine windings of all defects. As defects are 
far separated here, one can also determine the winding of each of the 
defect by noticing rotation of the brushes directly. The results are in 
complete agreement with the windings determined using our technique of
quadrant assignment. This is not surprising as the arguments given 
above for the technique use only continuity properties of the director, 
and hence are topological in nature, independent of details of the defect 
network.

 Note that whenever two (or more) brushes join defects, they represent 
defect-antidefect pairs. [It is easy to see that in between two defects 
of same windings there must be a region belonging to same quadrant 
extending to infinity, unless truncated by other defects \cite{sctr}.] 
Figs.1,2 show that defect-antidefect pairs are most 
abundant, supporting the correlation in defect-antidefect production.

 Pictures like in Fig.2 are present in the literature \cite{chndr}
(though we have not seen pictures as in Fig.1 with very dense network
of defects). However in some of those cases, one also observes few 
strength 1/2 defects, i.e.  points from which only two brushes emanate. 
We do not get any of these. If we had missed any such points due to 
resolution of the picture, it would have led to conflict in director 
assignment on the two sides of the brush following our technique, as we 
have verified from the pictures in the literature. A possible
explanation for the absence of 1/2 defects could be that these are 
point monopole defects. However, it is known that monopole production
in this manner is highly suppressed \cite{mnpl}. Further, similar
pictures in the literature \cite{chndr} show some strength 1/2 defects 
as well, which is not possible if these are monopoles. We
propose the following explanation for the absence of strength 1/2
defects in these pictures. The anchoring of the director at the 
I-N interface \cite{angle} forces the director to lie on a cone, with 
the half angle equal to about 64$^0$. This forces the order parameter 
space there to become effectively a circle $S^1$, instead of being 
$RP^2$, with the order parameter being an angle between 0 and $2\pi$. 
Only defects allowed now are with integer winding and no 1/2 windings 
can occur. Of course, depending on the anchoring energy, strength 1/2 
defects could still form, with certain region having higher energy. 

 Further, the space here is effectively two dimensional since integer 
windings can be trivialized as one moves away from the I-N interface, 
towards the nematic-air interface with normal boundary condition. (In 
this sense, these defects may be like partial monopole configurations.) 
Therefore the prediction for $\sigma$ from the Kibble mechanism for the 
U(1) case, as described above, is valid for this case, with the 
picture that a domain structure near the I-N interface is responsible
for the formation of integer windings.

 After identifying windings of all defects (wherever possible) in
a picture, we first determine the average defect density, and then 
divide the picture in terms of square shaped regions ($\Omega$) 
containing $N$ defects on average. We do the analysis for 
three different values of N, N = 10, 20 and 30.  
Regions are marked without noticing presence of defects to avoid any 
bias. In order to increase statistics, we also included some 
square regions with partial overlap (making sure that the boundaries of 
the two regions, though intersecting, should not overlap). It should be 
clear that net windings along the perimeters of such squares also 
represent independent statistics. For each square region, net defect 
number $\bigtriangleup n$ (i.e. number of defects minus number of 
antidefects), was found and by analyzing a large number of pictures, the 
frequency $f(\bigtriangleup n)$ of a given value of $\bigtriangleup n$ 
was determined.

 Fig.3 shows the plots of $f(\bigtriangleup n)$ vs.  $\bigtriangleup n$.
Solid, dotted and dashed curves show Gaussian fits to experimental points 
corresponding to N = 10, 20, and 30 respectively. Number of regions 
analyzed for these cases was 91, 54 and 34 in that order. Table 1 
summarizes our results for the Gaussian fits for 
the three sets of data, where we give the best fit values of the
parameters of the Gaussian, along with the standard errors in the 
determination of these parameters from the fit.

 For square shaped elementary domains, predicted values of $\sigma$ are 
1.26, 1.50, and 1.66, for N = 10, 20, and 30 respectively which are
in reasonable agreement with the measured values given in Table 1.
(For triangular domains, predicted values are lower, with 
$\sigma =$ 1.01, 1.21, and 1.33 for N = 10, 20, and 30 respectively.)
If defects and antidefects were uncorrelated then we expect, by randomly 
distributing defects and antidefects in the region,
that  $\sigma$ = 3.15, 4.47, and 5.50, for N = 10, 20 and 30 respectively. 
These values are markedly different from the values experimentally 
observed. This clearly rules out any mechanism of defect production where 
defects and antidefects are uncorrelated.  Note that if defect-antidefect 
pairs were thermally produced, then any resulting correlation could only 
be observed for typical inter-defect separations of order of the core 
size of the defect ($\simeq$ few hundred angstroms).  The inter-defect 
separation we observe (at the time of formation itself) is of the order 
of 30-40 microns. Though we mention here that some times we observe 
defects after the network has undergone some evolution. However, it does 
not affect correlations between defects and antidefects produced via the 
kibble mechanism (as long as kinetic energies of defects are not too 
large, which certainly is the case here).  Defect-antidefect 
symmetry implies that the Gaussians should be centered at zero. As we 
see from Table 1, centers of Gaussians $\overline{\bigtriangleup n}$
are indeed consistent with zero.

 Given the values of $\sigma$ for different N, we can determine the 
exponent $\nu$ in Eqn.(1). Stars in Fig.4 denote experimental values 
of $ln(\sigma)$ vs. $ln(N)$ for the three values of N. Straight
line shows the best linear fit to these points. The slope of the line
gives the value of the exponent $\nu$. We find,

\begin{equation}
\nu =  0.26 ~\pm ~ 0.11
\end{equation}

 This value is in excellent agreement with the theoretical value of 1/4 
predicted by the Kibble mechanism. Though error is somewhat large, it 
still rules out zero correlation between defects and antidefects which 
would give value of exponent to be 1/2. The intercept of the line in 
Fig.4 is found to be $- 0.27 \pm 0.27$. This gives the value of
the prefactor in Eqn.(1) to be $C =  0.76 \pm 0.21$. Again, this value 
is in good agreement with the predicted value of $C$ = 0.71 from the 
Kibble mechanism for the case of square shaped elementary domains,
though error is too large in this case for making any definitive 
statement about preferred shape of elementary domains. 

 We conclude by stressing that these observations provide first
measurement of defect-antidefect correlations, and lead to experimental 
verification of a crucial aspect of the Kibble mechanism. Another
point is that the prediction of defect density, via Kibble mechanism,
crucially requires the knowledge of the domain size \cite{zrk}. In 
ref.\onlinecite{lc2}, the transition proceeded by bubble nucleations, 
so domains were easily identified. When domains are not that clearly 
identifiable, as in the present case, then how does one determine the 
process underlying the defect production? Here, by checking a
qualitatively different aspect of the Kibble mechanism, one is able to
say that the correlations in defect-antidefect production support
the underlying picture being that of the Kibble mechanism.
We also emphasize that the technique we have described for determining 
windings of defects is an extremely efficient one, (and also fun 
to play with). We believe that this technique can be very useful in 
determining properties of dense defect networks in liquid crystals. 

 We are very thankful to Supratim Sengupta for useful discussions 
and comments. We would like to acknowledge V.S. Ramamurthy for 
his encouragement and help in setting up liquid crystal 
experiments at IOP.



\vskip 1in

\begin{table}[h]
\caption{Results of fitting data to $f(\bigtriangleup n) = a 
e^{-{(\bigtriangleup n - {\overline{\bigtriangleup n}})^2 \over 2 
\sigma^2}}$}
\begin{tabular}{llll}
N & a & ${\overline{\bigtriangleup n}}$ & $\sigma$ \\ \hline
10 & 26.37 $\pm$ 1.15 & 0.06 $\pm$ 0.07 & 1.41 $\pm$ 0.07 \\
20 & 13.54 $\pm$ 1.12 & -0.17 $\pm$ 0.16 & 1.64 $\pm$ 0.16 \\
30 & 7.21 $\pm$ 0.78 & 0.44 $\pm$ 0.24 & 1.94 $\pm$ 0.25 \\
\end{tabular}
\end{table}


\vskip 1in

\centerline{\bf Figure Captions}
\vskip .1in

1) Picture of defect distribution observed using cross-polarizers
in I-N transition. Size of the image is about 0.43 mm $\times$ 0.40 mm.

2) Verification of the procedure for identifying the windings of defects.

3) Plots of the frequency $f(\bigtriangleup n)$ vs. $\bigtriangleup n$.

4) Determination of the exponent $\nu$.

\newpage

\begin{figure}[h]
\begin{center}
\leavevmode
\epsfysize=13truecm \vbox{\epsfbox{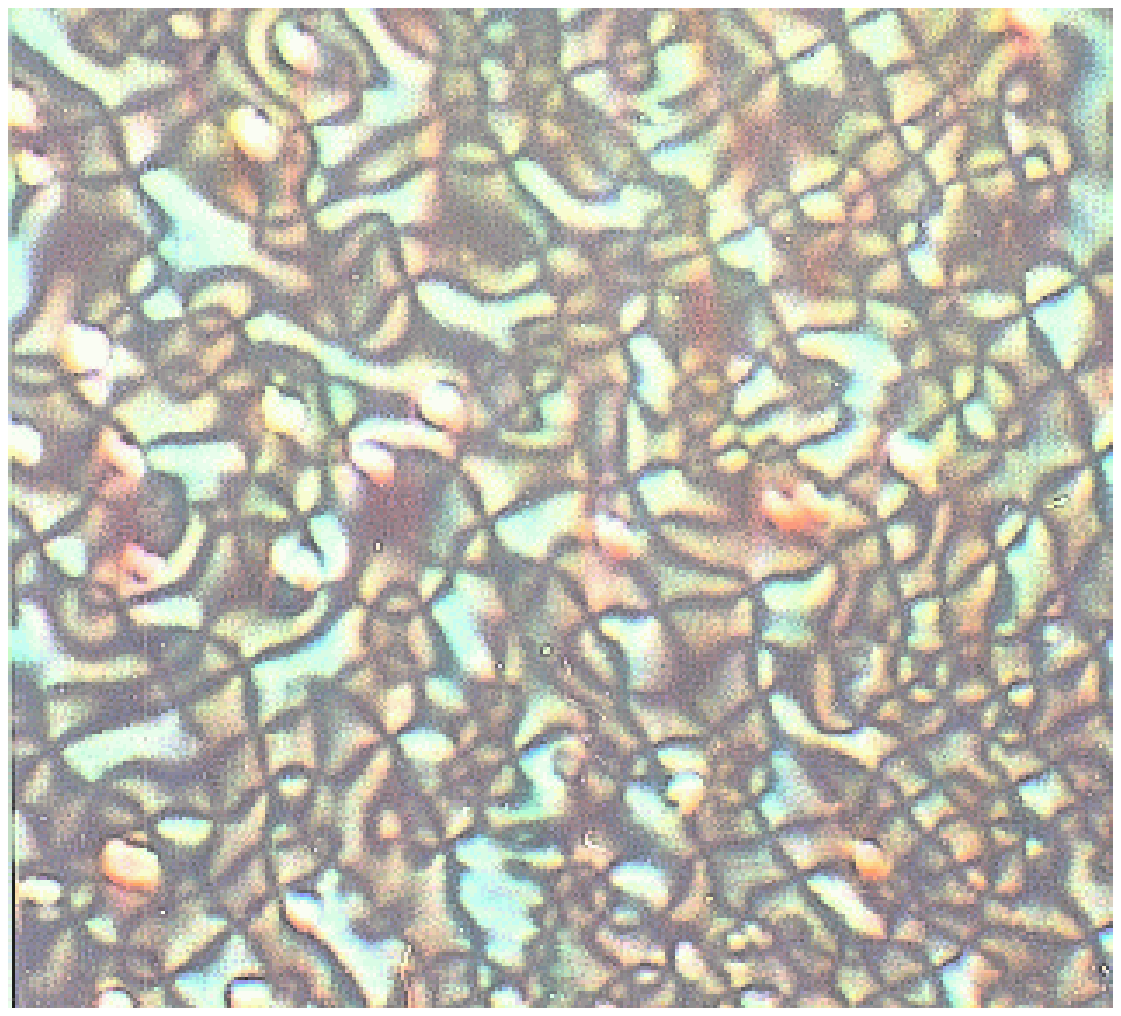}}
\end{center}
\caption{}
\label{Fig.1} 
\end{figure}

\newpage

\begin{figure}[h]
\begin{center}
\leavevmode
\epsfysize=8truecm \vbox{\epsfbox{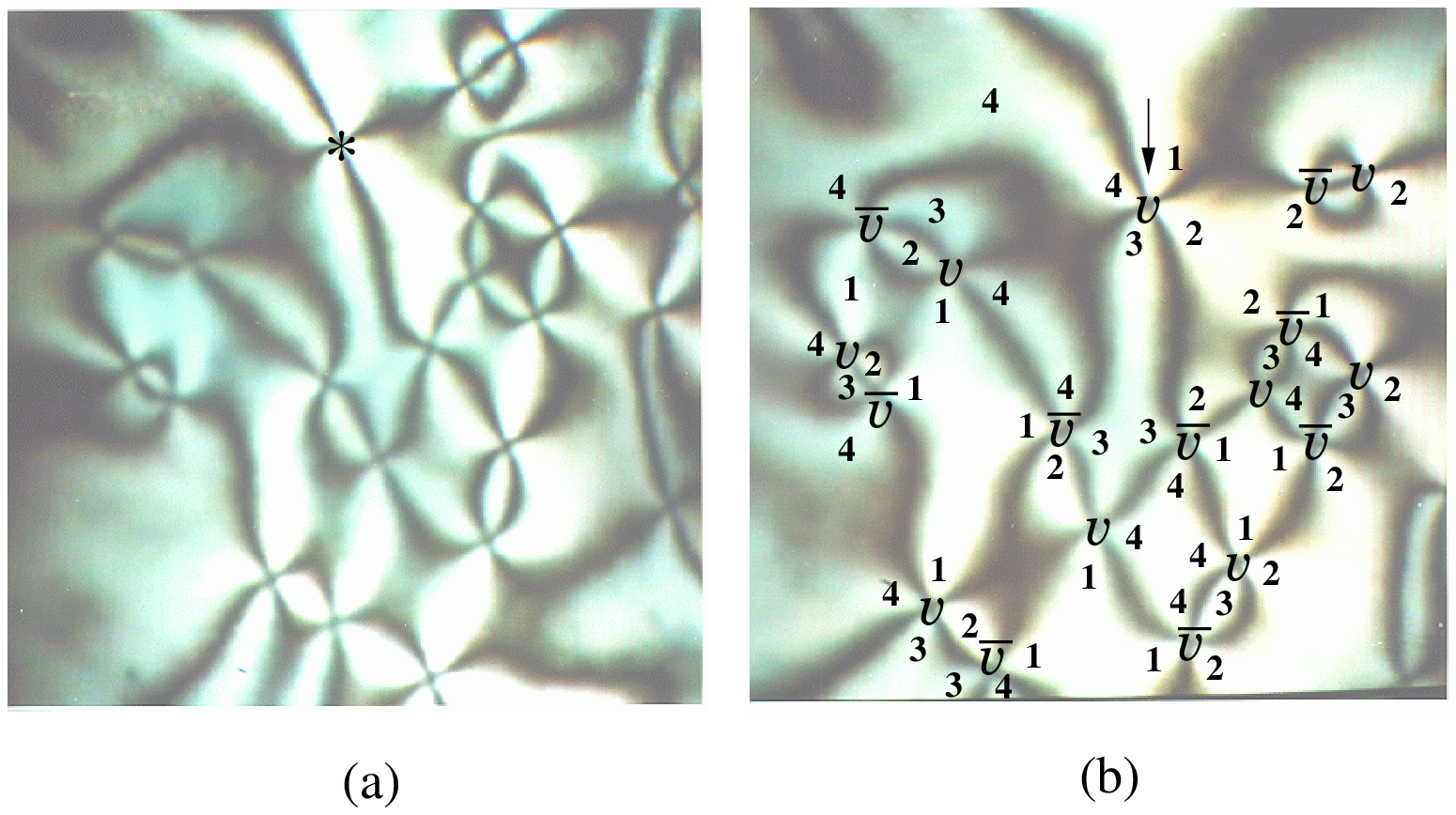}}
\end{center}
\caption{}
\label{Fig.2} 
\end{figure}

\newpage

\begin{figure}[h]
\begin{center}
\leavevmode
\epsfysize=17truecm \vbox{\epsfbox{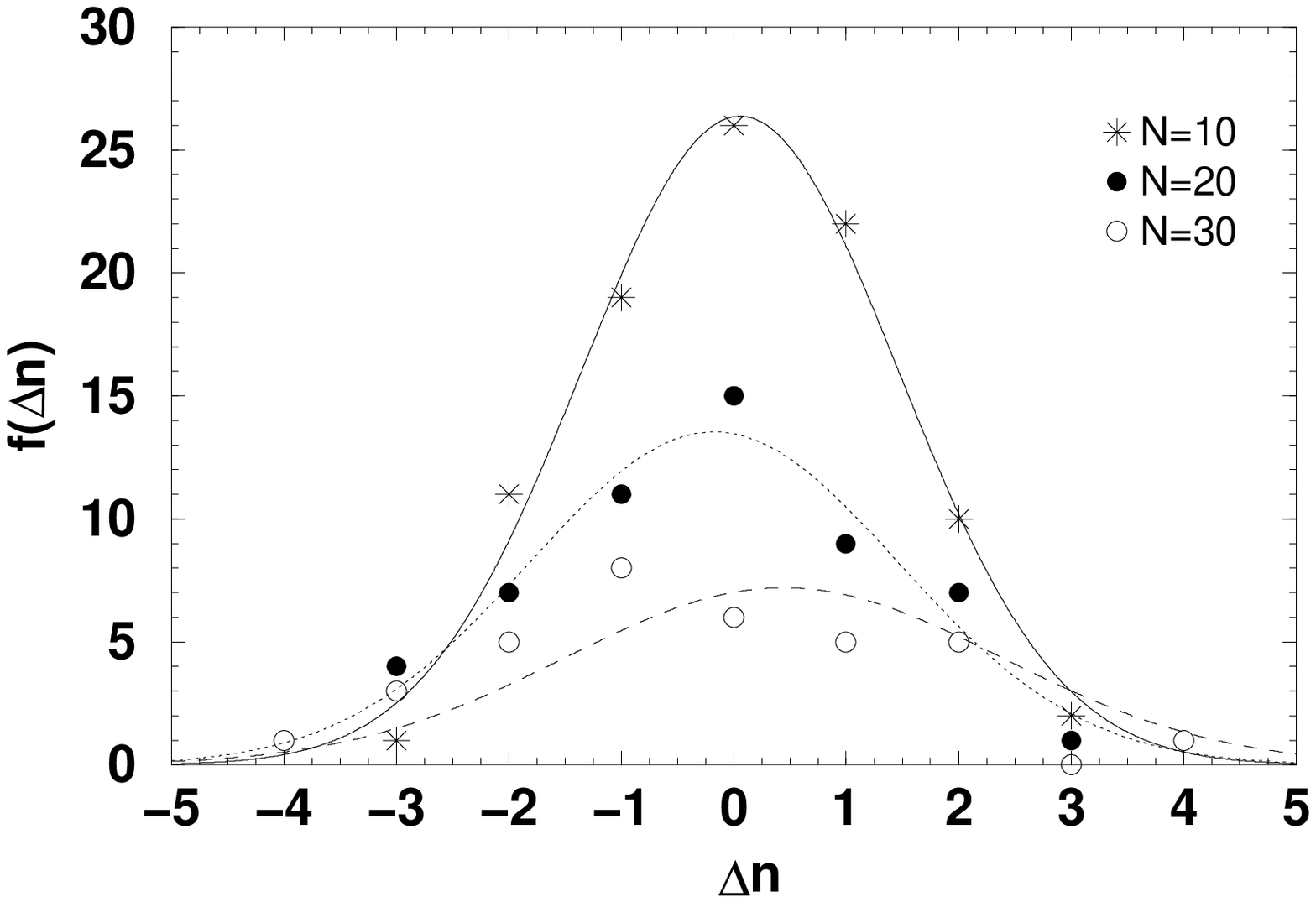}}
\end{center}
\caption{}
\label{Fig.3} 
\end{figure}

\newpage

\begin{figure}[h]
\begin{center}
\leavevmode
\epsfysize=17truecm \vbox{\epsfbox{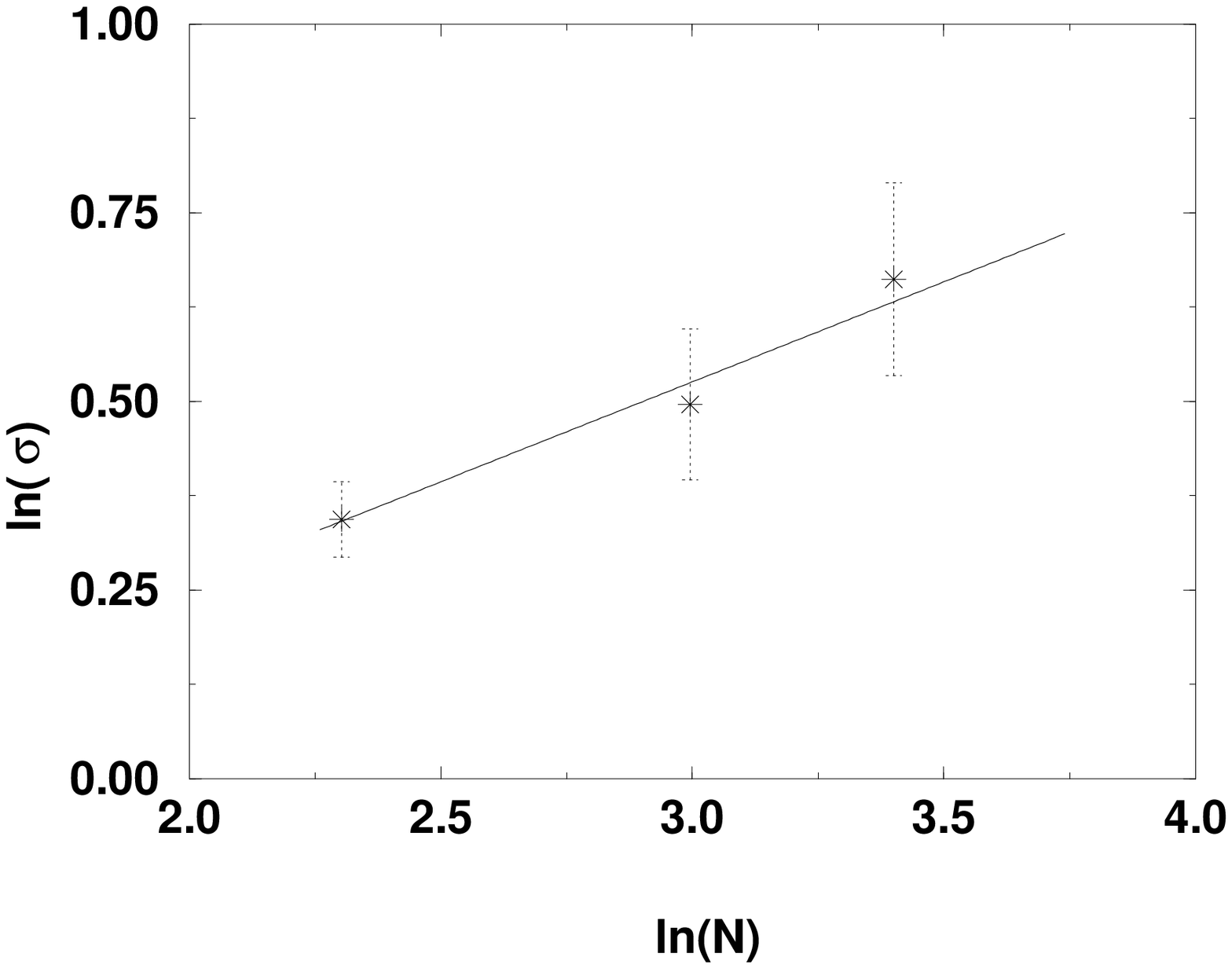}}
\end{center}
\caption{}
\label{Fig.4} 
\end{figure}

\end{document}